\documentclass[12pt]{article}
\usepackage{bm}
\usepackage{color}
\usepackage{dcolumn}
\usepackage{graphics}
\usepackage{graphicx}
\usepackage{subfigure}
\usepackage{slashed}
\usepackage{placeins}

\usepackage{pdfpages}
\usepackage{eso-pic}
\usepackage{everyshi}
\usepackage{pdflscape}

\newcommand{\ep}{\varepsilon}

\def\be{\begin{equation}}
\def\ee{\end{equation}}
\def\bea{\begin{eqnarray}}
\def\eea{\end{eqnarray}}
\def\bse{\begin{subequations}}
\def\ese{\end{subequations}}
\def\bc{\begin{center}}
\def\ec{\end{center}}

\def\I{{\rm i}}

\begin{document}

\begin{center}
  {\bf 
    Landau-Khalatnikov-Fradkin Transformation and Even $\zeta$ Functions  }\\

  \vskip 0.5cm
A.\ V.~Kotikov$^{1}$ and S.~Teber$^{2}$\\
  

 \vskip 0.5cm

 $^1$Bogoliubov Laboratory of Theoretical Physics, Joint Institute for Nuclear Research, 141980 Dubna, Russia.\\
$^2$Sorbonne Universit\'e, CNRS, Laboratoire de Physique Th\'eorique et Hautes Energies, LPTHE, F-75005 Paris, France.
\end{center}
  

\begin{abstract}

An exact formula that relates standard $\zeta$ functions and so-called hatted $\zeta$ ($\hat{\zeta}$) functions in all
orders of perturbation theory is presented. This formula is based on the Landau-Khalatnikov-Fradkin
transformation.
  
\end{abstract}


\section{Introduction }

We consider properties of multiloop massless functions of the propagator type. There is an ever
growing number of indications (see, for example, \cite{Baikov:2018gap})
that, in the calculations of various quantities in the
Euclidean region, striking regularities arise in terms
proportional to $\zeta_{2n}$--that is, to even Euler $\zeta$ functions.
These regularities are thought  \cite{Broadhurst:1999xk} to be due to the
fact that $\ep$-dependent combinations of $\zeta$ functions,
such as
\be
\hat{\zeta}_3 \equiv \zeta_3 + \frac{3\ep}{2} \zeta_4 - \frac{5\ep^3}{2} \zeta_6,~~ \hat{\zeta}_5 \equiv \zeta_5 + \frac{5\ep}{2} \zeta_6,~~
\hat{\zeta}_7 \equiv \zeta_7\, ,
\label{hatZe}
\ee
rather than the $\zeta$ functions themselves are dominant
objects that eliminate $\zeta_{2n}$ in the  $\ep$ expansions of four-loop functions belonging to the propagator type. A
generalization of combinations in (\ref{hatZe}) to the cases of
five, six, and seven loops can be found in \cite{Georgoudis:2018olj}.
\footnote{We note that the results in \cite{Georgoudis:2018olj} also contain multiple $\zeta$ functions (multi-zeta values),
  but their analysis is beyond the scope of the present article.}
  The
results in (\ref{hatZe}) and their generalization in \cite{Georgoudis:2018olj} make
it possible to predict $\pi^{2n}$ terms in higher orders of
perturbation theory.

In \cite{Kotikov:2019bqo} (see also \cite{Kotikov:2019gqp}), the present authors extended
the results in (\ref{hatZe}) to any order in  $\ep$ in a rather unexpected way---by means of the Landau-Khalatnikov-Fradkin
(LKF) transformation \cite{Landau:1955zz}, which relates the
fermion propagators in quantum electrodynamics
(QED) in two different gauges. It should be noted
that the most important applications of the LKF
transformation are generally associated with the predictions of some terms in high orders of perturbation
theory in QED \cite{Bashir:2002sp}, its generalizations \cite{Ahmad:2016dsb}, and more
general SU(N) gauge theories.

In the present article, we give a brief survey of the
results reported in \cite{Kotikov:2019bqo}, placing emphasis on how the
LKF transformation demonstrates in a natural way the existence of $\hat{\zeta}$ functions and makes it possible to
extend the results in (\ref{hatZe}) to any order in $\ep$.

\section{LKF transformation}
\label{sec:LKF:x-space}

Let us consider QED in $d$-dimensional ($d=4-2\ep$) Euclidean space. In general, the fermion propagator in a gauge
involving the parameter $\xi$  in the $p$
and $x$ representations has the form
\be
S_F(p,\xi) = \frac{1}{i\hat{p}} \, P(p,\xi) \, ,~~ S_F(x,\xi) =  \hat{x} \, X(x,\xi) \, ,
\label{SFp}
\ee
where there are explicit expressions for the factors $\hat{p}$ and $\hat{x}$, which involve the Dirac $\gamma$ matrices.

Within the dimensional regularization, the LKF transformation relates the fermion propagator in
these two gauges with parameters  $\xi$ and $\eta$, respectively, as \cite{Kotikov:2019bqo}
\be
S_F(x,\xi) = S_F(x,\eta)\, e^{\I D(x)} \, ,
\label{LKFN}
\ee
where
\be
D(x)=\frac{\I\,\Delta\,A}{\ep}\,\Gamma(1-\ep)\, (\pi\mu^2 x^2)^{\ep} ,~~ \Delta=\xi - \eta,~~
A=\frac{\alpha_{\rm em}}{4\pi}=\frac{e^2}{(4\pi)^2}\, .
\label{DxN}
\ee
This means that $D(x)$ makes a contribution proportional to $\Delta A$ and the pole  $\ep^{-1}$.

Suppose that, for a gauge-fixing parameter $\eta$, the
fermion propagator  $S_F(p,\eta)$ with an external momentum $p$ has the form (\ref{SFp}), where
\be
P(p,\eta) = \sum_{m=0}^{\infty} a_m(\eta)\, A^m \,{\left(\frac{\tilde{\mu}^2}{p^2}\right)}^{m\ep} \, ,~~ \tilde{\mu}^2= 4\pi \mu^2 \, .
\label{Peta}
\ee
Here, $a_m(\eta)$ are the coefficients in the loop expansion of the propagator and
$\tilde{\mu}$ is the renormalization scale lying between the scales of the ${\rm MS}$ (minimal-subtraction) 
and $\overline{{\rm MS}}$ (modified-minimal-subtraction)
schemes. The LKF transformation determines the
fermion propagator for another gauge parameter $\xi$ as
\be
P(p,\xi) = \sum_{m=0}^{\infty} a_m(\eta)\, A^m\, {\left(\frac{\tilde{\mu}^2}{p^2}\right)}^{m\ep}
\sum_{l=0}^{\infty} \, \frac{1-(m+1)\ep}{1-(m+l+1)\ep} \,
\Phi_{\rm MV}(m,l,\ep) 
\, \frac{(\Delta \, A)^l}{(-\ep)^l l!} \, {\left(\frac{\mu_{\rm MV}^2}{p^2}\right)}^{l\ep},
\label{axi.1}
\ee
where
%
\be
\Phi_{{\rm MV}}(m,l,\ep)=\frac{\Gamma(1-(m+1)\ep)\Gamma(1+(m+l)\ep)\Gamma^{2l}(1-\ep)}{
 \Gamma(1+m\ep)\Gamma(1-(m+l+1)\ep)} \, .
\label{Phi:V:def}
\ee
Here, the symbol ${\rm MV}$ stands for the so-called minimal Vladimirov scale introduced in \cite{Kotikov:2019bqo}. We note that,
in \cite{Kotikov:2019bqo}, the use of the popular G scale \cite{Chetyrkin:1980pr} led to the
same final results given in Eqs.~(16) and (17) below.

In order to derive expression (\ref{axi.1}), we employed
the fermion propagator $S_F(p,\eta)$ with  $P(p,\eta)$ given by (\ref{Peta}), applied the Fourier transformation to
$S_F(x,\eta)$, and made the LKF transformation (\ref{LKFN}). As a final step,
we performed the inverse Fourier transformation and
obtained the fermion propagator $S_F(p,\xi)$ with  $P(p,\xi)$ given in (\ref{axi.1}).

Let us now study the factor $\Phi_{{\rm MV}}(m,l,\ep)$. For this, we make use of the expansion of the $\Gamma$ function in the
form
\be
\Gamma(1+\beta\ep) = \exp \Big[ -\gamma \beta \ep + \sum_{s=2}^{\infty}\, (-1)^s \, \eta_s \beta^s \ep^s \Bigr],~~ 
\eta_s = \frac{\zeta_s}{s} \, ,
\label{Gamma:exp}
\ee
where $\gamma$ is the Euler constant.
Substituting this expansion into expression (\ref{Phi:V:def}), we recast the factor
$\Phi_{{\rm MV}}(m,l,\ep)$ into the form
\be
\Phi_{{\rm MV}}(m,l,\ep)= \exp \Big[ \sum_{s=2}^{\infty}\,\eta_s \, p_s(m,l) \, \ep^s \Bigr]\, ,
\label{Phi:V}
\ee
where
\be
p_s(m,l)=  (m+1)^s-(m+l+1)^s + 2l + 
(-1)^s \Bigl\{(m+l)^s-m^s\Bigr\},~~ p_1(m,l)=0,~~ p_2(m,l)=0 \, .
\label{ps:V}
\ee

One can readily see from Eq.~(\ref{Phi:V}) that the factor $\Phi_{{\rm MV}}(m,l,\ep)$
involves values of the $\zeta_s$ function of given weight $s$ (or transcendental level) in front of $\ep^s$.
This property constrains strongly the coefficients, thereby simplifying the ensuing analysis (the authors
of the articles quoted in \cite{Kotikov:2000pm} also used this property).

\section{$\hat{\zeta}_{2n-1}$
}
\label{sec:LKF:proof}

We now focus on the polynomial $p_s(m,l)$ in Eq.~(\ref{ps:V}). It is convenient to partition it into components
featuring even and odd values of s. The following recursion relations hold:
\be
	p_{2k} = p_{2k-1} + L p_{2k-2} + p_{3}, ~~ p_{2k-1} = p_{2k-2} + L p_{2k-3} + p_{3}, ~~  L=l(l+1) \, . 
\label{p2k-1:V}
\ee

Expressing even components, $p_{2k}$, in terms of odd ones as 
\be
p_{2k}=\sum_{s=2}^{k} p_{2s-1} \, C_{2k,2s-1} \, =  \sum_{m=1}^{k-1} p_{2k-2m+1} \, C_{2k,2k-2m+1} \, 
\label{p2k:V:general}
\ee
we can determine the exact structure of $C_{2k,2k-2m+1}$ in the form
\be
C_{2k,2k-2m+1} = b_{2m-1}
\, \frac{(2k)!}{(2m-1)! \, (2k-2m+1)!},~~
b_{2m-1} = \frac{(2^{2m} - 1)}{m} \, B_{2m} \, ,
\label{b:V:expressionB}
\ee
where $B_m$ are well-known Bernoulli numbers.

It is now convenient to represent the argument of
the exponential form on the right-hand side of Eq.~(\ref{Phi:V}) in the form
\be
\sum_{s=3}^{\infty}\,\eta_s \, p_s \, \ep^s = \sum_{k=2}^{\infty}\,\eta_{2k} \, p_{2k} \, \ep^{2k} +
\sum_{k=2}^{\infty}\,\eta_{2k-1} \, p_{2k-1} \, \ep^{2k-1} \, .
\label{Phi:V:exp}
\ee
With the aid of Eq.~(\ref{p2k:V:general}), the first term on the right-hand side of (\ref{Phi:V:exp}) can be represented in the form
\bea
\sum_{k=2}^{\infty}\,\eta_{2k} \, p_{2k} \, \ep^{2k} = \sum_{k=2}^{\infty}\,\eta_{2k}  \, \ep^{2k} \, \sum_{s=2}^k p_{2s-1} \,
C_{2k,2s-1} = \sum_{s=2}^{\infty} p_{2s-1} \, \sum_{k=s}^{\infty}\,\eta_{2k} \, C_{2k,2s-1} \, \ep^{2k} \, .
\nonumber
\eea

Relation~(\ref{Phi:V:exp}) can then be recast into the form
\be
\sum_{s=2}^{\infty}\,\hat{\eta}_{2s-1} \, p_{2s-1} \, \ep^{2s-1}=
\sum_{s=2}^{\infty}\,[\hat{\zeta}_{2s-1}/(2s-1)] \, p_{2s-1} \, \ep^{2s-1} \, ,
\label{Phi:V:exp2}
\ee
where
\be
\hat{\zeta}_{2s-1} = \zeta_{2s-1} +  \sum_{k=s}^{\infty}\,\zeta_{2k} \, \hat{C}_{2k,2s-1} \, \ep^{2(k-s)+1} 
\label{hZeta}
\ee
with
\be
\hat{C}_{2k,2s-1} = \frac{2s-1}{2k} \, C_{2k,2s-1}  
= b_{2k-2s+1} \, \frac{(2k-1)!}{(2s-2)! \, (2k-2s+1)!} \, .
\label{hC}
\ee
Relations (\ref{hZeta}), (\ref{hC}), and (\ref{b:V:expressionB}) lead to an expression for $\hat{\zeta}_{2s-1}$
in terms of standard $\zeta$ functions that is valid in
all orders of the expansion in $\ep$.

\section{Conclusions}
\label{sec:summary}

The recursion relations in (\ref{p2k-1:V}) between the even and odd components of the polynomial associated
with the factor  $\Phi_{{\rm MV}}(m,l,\ep)$ (\ref{Phi:V:def}) have been deduced
from the result in (\ref{axi.1}) obtained by means of the LKF transformation for the fermion propagator. These
recursion relations make it possible to express all results for the factor $\Phi_{{\rm MV}}(m,l,\ep)$
in terms of $\hat{\zeta}_{2s-1}$.
Expressions (\ref{hZeta}) and (\ref{hC}) for them are valid in any
order of perturbation theory.\\

A.V. Kotikov is grateful to the Organizing Committee of the Session-Conference of Nuclear Physics
Section at the Department of Physical Sciences, Russian Academy of Sciences, for the invitation.

\end{document}